\documentclass[journal=jacsat, manuscript=article]{achemso}

\usepackage[T1]{fontenc}       
\usepackage{lmodern} 
\usepackage{amsmath, amsthm, amscd, amssymb} 
\usepackage[utf8]{inputenc} 
\usepackage[T1]{fontenc} 
\usepackage{graphicx} 
\usepackage{setspace}
\usepackage[hypertexnames=false]{hyperref} 
\usepackage{natbib} 
\usepackage{achemso}
\def \dif {\mathrm{d}}

\author{Gerard P. Conangla}
\email{gerard.planes@icfo.eu}
\affiliation{ICFO Institut de Ciencies Fotoniques, Mediterranean Technology Park, \linebreak 08860 Castelldefels (Barcelona), Spain}

\SectionNumbersOn
\title{Sensing with the harmonic oscillator}

\begin{document}

\begin{abstract} 
A system obeying the harmonic oscillator equation of motion can be used as a force or proper acceleration sensor. In this short review we derive analytical expressions for the sensitivity of such sensors in a range of different situations, considering noise of thermal and measurement origins and a formalism for dealing with oscillators whose natural frequency $\omega_0$ jitters. A special case where the sensitivity can be improved beyond the standard expressions and some applications with examples are also discussed.
\end{abstract}

\section{Introduction}

Linear force and acceleration sensors are often based on systems whose equation of motion is well approximated by a harmonic oscillator. The inertial sensing market is nowadays dominated by micro-electromechanical systems (MEMS)\citep{shaeffer2013mems}, devices with sizes ranging between 20 $\mu$m to a mm that can be found in most modern cellphones\footnote{In fact, smartphones contain Inertial Measurement Units (IMU), devices incorporating both MEMS Accelerometers and MEMS gyroscopes, which provide linear acceleration and torque measurements respectively.}, but any \emph{harmonic oscillator} sensor follows the same working principles. For instance, a massive pendulum around its stable equilibrium position can be used if a large oscillator mass is convenient; if, on the contrary, the mass is required to be small, a levitated nano-particle or a micro-cantilever can be chosen.

Conceptually, force sensors and linear inertial sensors (also known as \emph{accelerometers}) detect the effect of a driving force as a displacement of the oscillator's mass; this displacement can be measured by various methods\footnote{For instance, modern MEMS accelerometers often consist of little more than a cantilever beam with a proof mass (also known as seismic mass), whose movement is measured as a time dependent capacitance. Other accelerometers work by detecting the electric field due to the strain applied to a piezoelectric crystal.}, depending on the sensor. In any case, the displacement is created when the driving modifies the equation of motion: the subtle difference is that in force sensors the driving only acts on the oscillator's mass, whereas in an accelerometer the driving acts on the oscillator's housing.

In this short review we derive analytical expressions for the sensitivity of such sensors in a range of different situations, considering noise of thermal and measurement origins and a formalism for dealing with oscillators whose natural frequency $\omega_0$ jitters. A special case where the sensitivity can be improved beyond the standard expressions and some applications with examples are also discussed.

\section{Force sensing}
\subsection{Principles of force sensing}
The harmonic oscillator is a 2nd order constant coefficient linear ordinary differential equation. In the most general case, the equation reads
\begin{align}
m\ddot{x} + m\Gamma \dot{x} + m\omega_0^2x = g(t)
\end{align}
where $m\Gamma\dot{x}$ is a damping force, $m\omega_0^2x$ is a restoring force and $g(t)$ is an (external) driving force. Force sensing is the detection of these forces $g(t)$ acting \emph{upon} the oscillator's mass. Since the harmonic oscillator is a linear time-invariant system (see \hyperref[LTI]{supplementary}), the response of the system to a deterministic (arbitrary) driving function $g(t)$ will be
\begin{align}
x(t) = h(t) * g(t),
\end{align}
where $h(t)$ is the impulse response of the system. By the convolution theorem, $X(\omega) = H(\omega)\cdot G(\omega)$, where the capital letters indicate Fourier transforms. Recall the transfer function of the harmonic oscillator (see \hyperref[harmonic_transfer]{supplementary}) is
$$
H(\omega) = \frac{1/m}{\omega_0^2 - \omega^2 + i \omega \Gamma^2}.
$$

If the driving function is stochastic, but wide sense stationary\footnote{This assumption is important, since the following expressions assume no transient behaviours. When $\Gamma \ll 1$, past transients can contaminate the signal, leading to worse signal to noise ratio (SNR) than expected.}, this equality still holds in the power spectral density (PSD) sense
$$
S_{xx}(\omega) = |H(\omega)|^2\cdot S_{gg}(\omega).
$$
Therefore, given a certain experimentally measurable $x(t)$, to obtain its originating $S_{gg}$ we would need to calculate
\begin{align}\label{eq:force_measurement}
S_{gg}(\omega) = |H(\omega)|^{-2}\cdot S_{xx}(\omega).
\end{align}

In a more realistic scenario, the oscillator will be driven by thermal noise\footnote{As can be seen by the fluctuation-dissipation theorem\citep{Kubo1966}. The engineer/scientist should decide whether this Brownian noise can be neglected or should be taken into account, depending on the harmonic oscillator's mass.}, a stochastic driving that can be modelled as a white noise $w(t)$ with zero mean and autocorrelation function $R_w(\tau) = \sigma_1^2 \delta(\tau)$. There will also be a random noise $u(t)$ originating from our measuring device, that we can consider additive Gaussian white noise (AGWN) (with $R_u(\tau) = \sigma_2^2 \delta(\tau)$), summed \emph{after} the harmonic oscillator frequency response. Therefore, the measured power spectral density of a driving force $g(t)$ is
$$
S_{xx} = |H(\omega)|^{2}S_{gg} + |H(\omega)|^{2}S_{ww} + S_{uu} =
|H(\omega)|^{2}S_{gg} + |H(\omega)|^{2}\sigma_1^2 + \sigma_2^2,
$$
and our estimate of $S_{gg}$ will be
\begin{align}
|H(\omega)|^{-2}\cdot S_{xx} & = \nonumber
|H(\omega)|^{-2}\left(|H(\omega)|^{2}S_{gg} +
|H(\omega)|^{2}\sigma_1^2 + \sigma_2^2\right)\\ 
& = S_{gg} + \sigma_1^2 + |H(\omega)|^{-2}\sigma_2^2.
\end{align}
In this expression $S_{gg}$ is deterministic, $\sigma_1^2$ comes from the thermal noise and is independent of $\omega$ and
$|H(\omega)|^{-2}\sigma_2^2$ has explicit $\omega$ dependency. The signal to noise ratio (SNR), defined as the square root\footnote{We use the square root to work with force units, instead of power units. This is, of course, arbitrary.} of the ratio of the signal and noise powers, will thus take the expression
$$
\text{SNR} = \sqrt{\frac{S_{gg}}{\sigma_1^2 + |H(\omega)|^{-2}\sigma_2^2}}.
$$
This expression is bounded by $\sqrt{\frac{S_{gg}}{\sigma_1^2}}$: this is a hard limit that cannot be improved with this system and assumptions. However, the $|H(\omega)|^{-2}\sigma_2^2$ term can be minimised by using $\omega$ at resonance.

In any case, to get the minimum measurable force $g(t)$ we have to set a limit to what we can detect: usually this limit is $\text{SNR} > 1$\footnote{Again, this limit is arbitrary: in principle if the system is ergodic and the driving force periodic, signals of any SNR can be detected}. From this inequality we can obtain a bound for $g(t)$, which is where the minimum sensitivity expressions come from. Using the fluctuation-dissipation relationship $\sigma_1 = \sqrt{2m\Gamma k_B T}$ and $|H(\omega)|^2$ at its maximum (i.e., at resonance) we get
\begin{align}\label{eq:minsensing}
\sqrt{S_{gg}} & > \sqrt{\sigma_1^2 + |H(\omega)|^{-2}\sigma_2^2} \nonumber \\
& = \sqrt{2m\Gamma k_B T + \sigma_2^2\frac{m^2(4Q^2-1)\omega_0^4}{4Q^4}} \simeq \sqrt{2m\Gamma k_B T + \sigma_2^2m^2\omega_0^2\Gamma^2}
\end{align}
where after the inequality we assumed resonance, and the last equality is a good approximation for large enough $Q$. Whether we are measuring at resonance or not, if the measuring noise is very small compared to the thermal noise, the second term in the square root may be negligible, in which case we obtain the following expression for the amplitude spectral density (ASD)
\begin{align} \label{approx_asd}
\text{ASD} = \sqrt{2 m \Gamma k_B T} 
\end{align} 
The ``minimum resolvable'' \citep{armano2016sub} force or force sensitivity is hence defined as the root mean square of the noise power spectral density (i.e., the ASD is just the square root of the PSD\footnote{For reference, check the LISA Pathfinder \href{https://journals.aps.org/prl/pdf/10.1103/PhysRevLett.116.231101}{paper}.}). Notice that \emph{smaller} sensitivity values are \emph{better}, since the benchmark is the smallest signal that can be detected. 

Expression \eqref{approx_asd} is useful to get an intuition of how the sensitivity scales with the oscillator parameters. For example, a smaller mass results in a better sensitivity\footnote{Assuming, of course, that noise has no $m$ dependency. For instance, with optically levitated nanoparticles, decreasing the mass increases the measurement noise, since smaller particles scatter less light. Thus, to obtain a signal that is comparable to that of bigger particles, we will need to amplify the measured signal, hence also amplifying measurement noise.}. Bear in mind, however, that \eqref{approx_asd} is just an approximation and the full sensitivity has $\omega$ dependency, 
\begin{align}\label{eq:full_sensitivity}
\text{ASD} = \sqrt{\sigma_1^2 + |H(\omega)|^{-2}\sigma_2^2}. \qquad (\text{N}/\sqrt{\text{Hz}})
\end{align}

If we further integrate the power only on an interval of length $\Delta f$ in the Fourier domain\footnote{As is done in ref. \citep{ranjit2015attonewton}.}, we obtain
$$
g_\text{min} = \sqrt{2 \Delta f S_{gg}} > \sqrt{4 m \Gamma \Delta f k_BT}.
$$
This expression of $g_\text{min}$ is also sometimes known as the force sensitivity of the system. Nevertheless, the previous ASD definition can (and is) also used under the same name. In general, the use of the force sensitivity defined as the ASD (and not as this $g_\text{min}$) should always be preferred, since its value doesn't depend on how long the measurement is, it doesn't make any assumptions on the bandwidth of the signal and any information of different sensitivities at different $\omega$ values is not lost.

It is worth pointing out that if one assumes that measurement noise can be neglected, the sensitivity is the same for all frequencies: it doesn't matter if one measures at resonance or out of it\footnote{In fact, measuring at resonance may be a bad idea if the driving signal has a big bandwidth, since the phase response changes a lot around the resonance peak.}. This happens because the oscillator gain affects thermal noise and driving signal equally and the SNR stays constant. Therefore, for a fixed $m$, the obvious knobs for reducing the ASD are decreasing the temperature $T$ and/or $\Gamma$.

\subsection{The case of a harmonic oscillator with jitter}

If the response function of the harmonic oscillator jitters (i.e., is not constant with time), we can model the sensitivity in probabilistic terms. Assume the jittering is due to a stochastic natural frequency of the oscillator $\Omega$\footnote{In contrast with the previous $\omega_0$.}, that now fluctuates with time, but has a certain stationary, well-known probability density function $f_\Omega(u)$. Then, the value of the ASD from \eqref{eq:full_sensitivity} will also be stochastic, but we can still calculate its statistical moments. In particular, the expected value of the sensitivity will be
\begin{align}
\mathbb{E}[\text{ASD}(\omega)] & = \mathbb{E}\left(\sqrt{\sigma_1^2 + |H(\omega)|^{-2}\sigma_2^2} \right) \\
& = \int_{-\infty}^\infty \sqrt{\sigma_1^2 + |H(\omega)|^{-2}\sigma_2^2} \cdot f_\Omega(u) \,\mathrm{d}u.
\end{align}
Clearly this value depends on $\omega$, since the expected sensitivity will not be the same for every frequency. However, if, as before, measurement noise can be neglected, then
\begin{align}
\mathbb{E}[\text{ASD}(\omega)] & = \int_{-\infty}^\infty \sqrt{\sigma_1^2 + |H(\omega)|^{-2}\sigma_2^2} \cdot f_\Omega(u) \,\mathrm{d}u \\
& \simeq \int_{-\infty}^\infty \sqrt{\sigma_1^2} \cdot f_\Omega(u) \,\mathrm{d}u. = \sigma_1,
\end{align}
recovering the expression of \eqref{approx_asd}. Intuitively, this means that if the natural frequency $\Omega$ jitters only in a region where thermal noise is still dominant (for instance the standard deviation of the natural frequency, $\sigma_\Omega$, is less than $\Gamma/2$), the sensitivity is not affected, even if the force amplitude fluctuates due to the jitter. \emph{However}, the actual value of the measured force will be affected if one naively uses equation \eqref{eq:force_measurement} without taking into account that $H(\omega)$ is stochastic. To calculate the real expected force, we will need to, again, find the expected value of $H(\omega)$
\begin{align}
\sqrt{S_{gg}(\omega)} = \sqrt{\mathbb{E}(|H(\omega)|^{-2})\cdot S_{xx}(\omega)},
\end{align}
which can be thought of as an effective response function obtained by averaging the instantaneous responses at different times.

Real oscillators, and especially small ones (which are more susceptible to perturbations), will always have some jitter. The importance of the effect depends on the parameters of the oscillator: for instance, the effect can be neglected when the damping is large and the oscillator's width is much broader than the jitter (i.e., $\Gamma \gg \sigma_\Omega$). However, for strongly underdamped systems -- which have a very narrow response peak --, jitter needs to be taken into account. Beyond the probabilistic treatment that we just described, it is advisable to have some way of artificially increasing the damping in the harmonic oscillator. As long as this extra damping does not induce additional noise (i.e., it is a \emph{cold damping}\citep{conangla2018optimal}), this has several advantages:
\begin{itemize}
\item Since the peak is broadened, the relative effect of jitter is reduced.
\item As long as thermal noise is still considerably larger than measurement noise, the SNR (i.e., the sensitivity) will not be affected. 
\item It will reduce the effect of non-linearities in the oscillator. Low dissipation systems with narrow resonances are prone to large oscillation amplitudes: in this case, non-linearities can dominate\footnote{\href{https://www.youtube.com/watch?v=j-zczJXSxnw}{This} is a classical example.}.
\item It will also reduce the memory of the system. Recall that, the lower the damping, the longer the memory of the impulse response. In practical terms, this means that if no additional damping is included we will keep measuring signal remnants from some past event long after the driving has stopped.
\end{itemize}

One possibility to implement this damping is with a feedback force $-mk_d\dot{x}$. The equation of motion becomes
\begin{align}
m\ddot{x} + m\Gamma \dot{x} + m\omega_0^2x = -mk_d \dot{x},
\end{align}
resulting in a new transfer function
$$
H(\omega) = \frac{1/m}{\omega_0^2 - \omega^2 + i \omega (\Gamma + k_d)^2},
$$
which has a bigger effective damping than before, while the thermal noise will still have spectral density $\sqrt{2k_B T m \Gamma}$. Real feedback systems will never be completely noise-free, but for small gains the effect of this noise is negligible (see supplementary material of ref. \citep{conangla2018optimal}).

\subsection{Sensing a perfectly sinusoidal force}
Assume now the force $g(t)$ we want to measure is a sinusoid, with a constant phase relation to a controlled reference signal (i.e., it is a \emph{perfect} sinusoid). In this case, the sensitivity to this signal can be greatly increased if instead of working with PSDs one averages the Fourier transform measurements (note that this gives $(\text{N}/\sqrt{\text{Hz}})$ units directly). The argument is described in what follows: the measured signal $m(t)$ will take the expression
$$
m(t) = (g(t) + w(t))*h(t) + u(t)
$$
where, as before, $g(t)$ is the driving force, $w(t)$ is a white thermal noise and $u(t)$ is an AWGN from the measuring device. By assumption, $g(t) = \cos w_d t$. Then, if we take the Fourier transform of the previous expression we get
\begin{align}
\mathcal{F}\left((g(t) + w(t))*h(t) + u(t)\right) = \mathcal{F}\left(g(t) * h(t)\right) + \mathcal{F}\left(w(t) * h(t)\right) + \mathcal{F}\left(u(t)\right) \\
= \pi (\delta(\omega - \omega_d) + \delta(\omega + \omega_d))\cdot H(\omega) + \nonumber \\
\mathcal{F} \left( \begin{pmatrix}
\cos \sqrt{a}t  \sin \sqrt{a}t/\sqrt{a}
\end{pmatrix}
\cdot 
\int_0^{t}
b e^{\frac{\Gamma}{2}(r-t)}\cdot
\begin{pmatrix}
-\sin \sqrt{a}r/\sqrt{a}\\ 
\cos \sqrt{a}r
\end{pmatrix}
\dif W_r \right) + 
\mathcal{F}\left(u(t)\right),
\end{align}
where we have used the properties of the Fourier transform, in the second summand we have substituted by the solution of a thermally driven harmonic oscillator (see the \hyperref[stochastic_solution]{supplementary material} for the derivation), $a = \omega_0^2 - \Gamma^2/4$ and $W_r$ is a Wiener process parametrized by the time $r$. Finally, taking expected values and applying Fubini's theorem
\begin{align}
\mathbb{E}[\mathcal{F}\left((g(t) + w(t))*h(t) + u(t)\right)] 
= \mathbb{E}[\pi (\delta(\omega - \omega_d) + \delta(\omega + \omega_d))\cdot H(\omega) + \\
\mathcal{F} \left( \begin{pmatrix}
\cos \sqrt{a}t  \sin \sqrt{a}t/\sqrt{a}
\end{pmatrix}
\cdot 
\int_0^{t}
b e^{\frac{\Gamma}{2}(r-t)}\cdot
\begin{pmatrix}
-\sin \sqrt{a}r/\sqrt{a} \nonumber \\ 
\cos \sqrt{a}r
\end{pmatrix}
\dif W_r \right) + 
\mathcal{F}\left(u(t)\right)] \nonumber \\ 
= \pi (\delta(\omega - \omega_d) + \delta(\omega + \omega_d))\cdot H(\omega) + \nonumber \\
\mathcal{F} \left( \begin{pmatrix}
\cos \sqrt{a}t  \sin \sqrt{a}t/\sqrt{a}
\end{pmatrix}
\cdot 
\mathbb{E}\left[
\int_0^{t}
b e^{\frac{\Gamma}{2}(r-t)}\cdot
\begin{pmatrix}
-\sin \sqrt{a}r/\sqrt{a} \nonumber \\ 
\cos \sqrt{a}r
\end{pmatrix}
\dif W_r \right)\right] + \mathcal{F}\left(\mathbb{E}[u(t)]\right) \\
= \pi (\delta(\omega - \omega_d) + \delta(\omega + \omega_d))\cdot H(\omega),
\end{align}
since the expected values of an Ito integral and $u(t)$ are zero. If we are perfectly rigorous, Fubini's theorem can't be applied with delta distributions; nonetheless, the calculation can be repeated in complete analogy with sinc functions instead. Sinc functions appear as the Fourier transforms of finite rectangular windows, and are unavoidable in actual measurements\footnote{Since every real measurement will be a \emph{finite} time measurement. Therefore, the measured signal of an arbitrary $f(t)$ will in fact be $f(t)\cdot \Pi(t)$ ($\Pi(t)$ being a rectangular window), and by the convolution theorem the Fourier transform of the measurement will be $F(\omega) * \text{sinc}(\omega)$. This avoids the infinite values from the delta distribution.}.

Thus, defining the SNR as the ratio between expected driving signal amplitude and expected noise amplitude in the frequency domain, we find that the value goes to infinity. A more accurate analysis should compare not the ratio of the expected values but the ratio of the signal and noise random variables themselves (which, to a good approximation, should follow a Cauchy distribution\footnote{The Cauchy distribution appears as the ratio of two normally distributed random variables and has undefined (i.e., going to infinity) statistical moments.}). However, this section is enough to show that if more information about the driving force is known (in this case, the fact that the force is sinusoidal), we don't need to restrict ourselves to the sensitivity described in eq. \eqref{eq:full_sensitivity} and more intelligent approaches might exist.

\section{Accelerometers}

In the last section we saw the expressions for the sensitivity and response of the harmonic oscillator when subjected to an external driving. In a situation where the driving acts on the housing of the oscillator instead of on the oscillator itself, the equation of motion for the harmonic oscillator is modified to
\begin{align}
m\ddot{x} + m\Gamma \dot{x} + m\omega_0^2x = 0 \rightarrow m\ddot{x} + m\Gamma (\dot{x} - \dot{y}) + m\omega_0^2(x - y) = 0
\end{align}
where $y(t)$ is the motion of the housing. By rewriting the equation in terms of $z \equiv x- y$, which is the quantity that we will measure, we obtain the equation
\begin{align}
m\ddot{z} + m\Gamma \dot{z} + m\omega_0^2z = -m\ddot{y}.
\end{align}

Therefore, the transfer function for $y(t)$, as compared with a regular harmonic oscillator, will take the modified expression
\begin{align}
H(\omega) = \frac{-\omega^2}{\omega_0^2 - \omega^2 + i \omega \Gamma^2}.
\end{align}

\begin{figure}
\begin{center}
\includegraphics[width=0.6\textwidth]{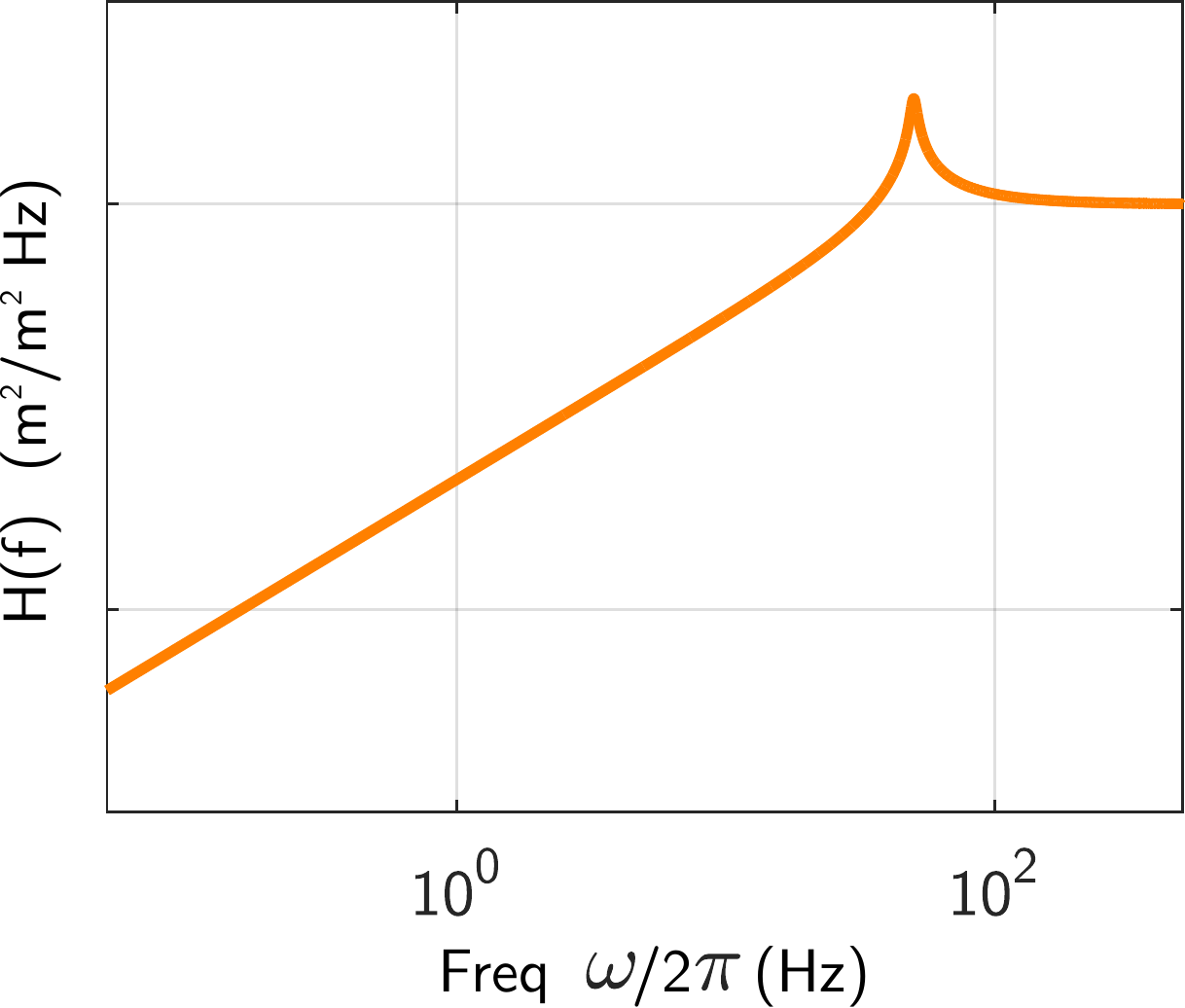}
\caption{Oscillator response $H(f)$, for arbitrary $\omega_0$ and $\Gamma$, to a housing motion $y(t)$ (seismometer regime).}
\label{fig:seismometer}
\end{center}
\end{figure}

Note that, for large values of $\omega$, $H(\omega) \simeq 1$. Therefore, for frequencies significantly above resonance, the accelerometer has a flat response and behaves approximately as a dirac delta:
$$
Z(\omega) = H(\omega)Y(\omega) =_{\omega \gg \omega_0} Y(\omega) 
$$
so
$$
z(t) \simeq y(t).
$$
In this range of operation, the oscillator behaves as a \emph{seismometer}: it measures the displacement of the housing (see Fig. \ref{fig:seismometer}).

To understand the \emph{accelerometer} regime, lets assume we have an arbitrary housing motion $y(t)$. The response in terms of the power spectral densities will be
\begin{align}
S_{zz}(\omega) = \frac{\omega^4}{(\omega_0^2 - \omega^2)^2 + \Gamma^2 \omega^2} \cdot S_{yy}(\omega).
\end{align}
But since $-\omega^2 Y(\omega) = A(\omega)$, where $Y(\omega)$ and $A(\omega)$ are the Fourier transforms of $y(t)$ and the acceleration $a(t) = \ddot{y}(t)$, the response to an acceleration $\ddot{y}(t)$ will be
\begin{align}
S_{zz}(\omega) = \frac{\omega^4}{(\omega_0^2 - \omega^2)^2 + \Gamma^2 \omega^2} \cdot S_{yy}(\omega) = \frac{1}{(\omega_0^2 - \omega^2)^2 + \Gamma^2 \omega^2} \cdot S_{aa}(\omega).
\end{align}

\begin{figure}
\begin{center}
\includegraphics[width=0.6\textwidth]{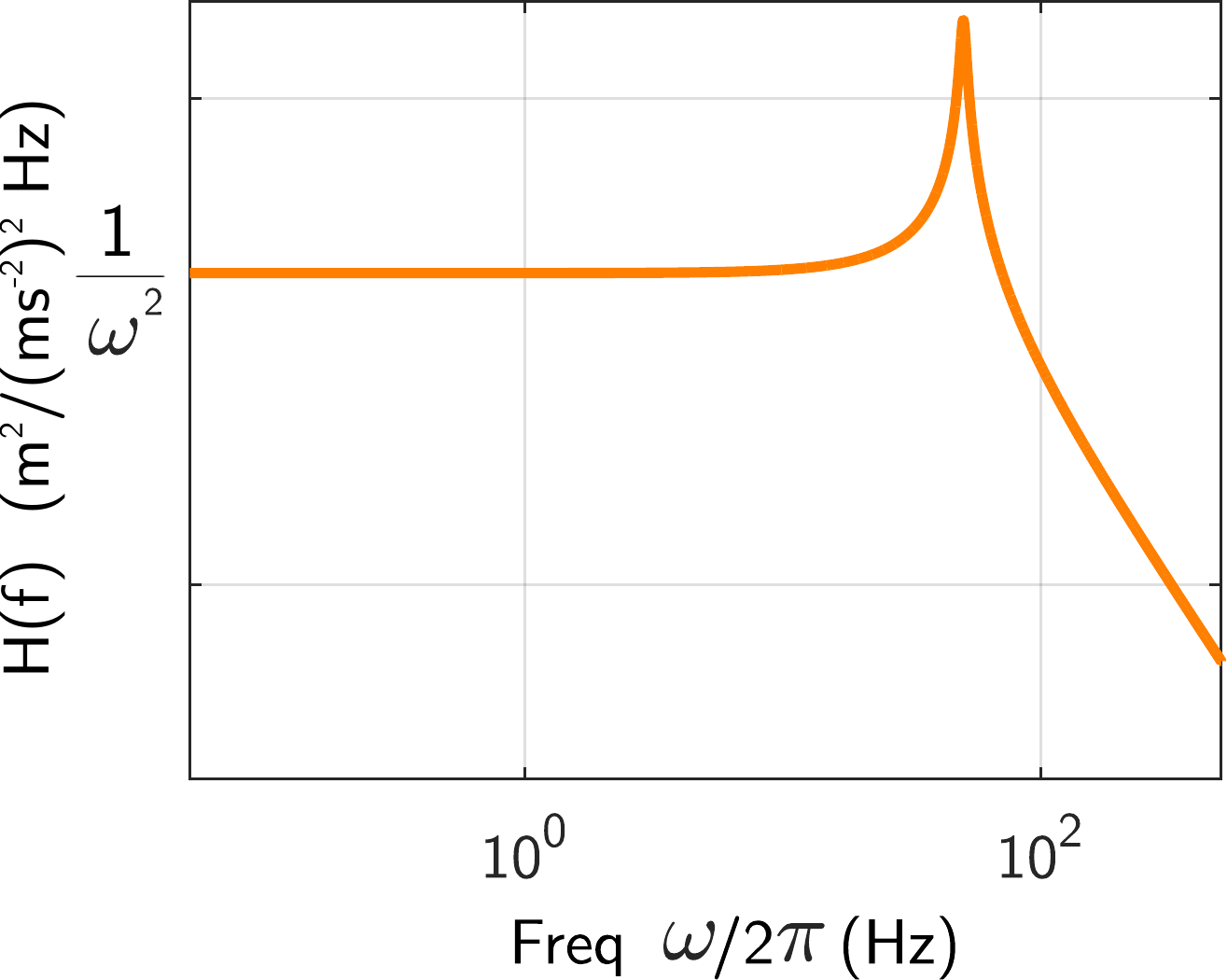}
\caption{Oscillator response $H(f)$, for arbitrary $\omega_0$ and $\Gamma$, to the acceleration of the housing $a(t)$ (accelerometer regime).}
\label{fig:accelerometer}
\end{center}
\end{figure}

In other words, the response of $z(t)$ to an acceleration of the housing is, except for the missing $1/m^2$ factor, the same as the response of a force acting on the harmonic oscillator, as plotted in Fig. \ref{fig:accelerometer}. Now, the response of the harmonic oscillator is flat at low frequencies (i.e., below resonance)
$$
H(\omega)_{\omega \rightarrow 0} = \frac{1}{\omega_0^2},
$$
and this is the frequency band where most of the accelerometers work. Like in seismometers, this is a relevant point: working in a regime where the impulse response is approximately a Dirac delta is necessary when the driving force isn't restricted to a small bandwidth (otherwise, to recover $a(t)$ out of $z(t)$, some sort of deconvolution may be needed, which is an ill-posed problem that should be avoided if possible).

Assuming the housing mass is large enough, the stochastic component of $y(t)$ due to Brownian noise can be neglected. However, Brownian noise will still drive the harmonic oscillator itself\footnote{Note that an important approximation is being done here: we assume that the acceleration leaves the statistical properties of Brownian noise unchanged. For large accelerations this approximation may not hold.}
$$
m\ddot{z} + m\Gamma \dot{z} + m\omega_0^2z = \omega(t).
$$
Therefore, repeating the analysis of the force sensing section we obtain
\begin{align}
S_{zz} = m^2 \cdot |H(\omega)|^2 S_{aa}  + |H(\omega)|^2\sigma_1^2,
\end{align}
where $H(\omega)$ is the response of the harmonic oscillator including the $1/m^2$ factor. Thus, neglecting the measurement noise, $\text{SNR} = \sqrt{\frac{S_{aa}m^2}{\sigma_1^2}}$ so
\begin{align}
\text{ASD} = \sqrt{S_{aa}} = \sqrt{\frac{2k_B T\Gamma}{m}}
\end{align}
This means that larger masses will push down the value of the minimum detectable accelerations.
%

\begin{acknowledgement}
The author acknowledges financial support from the European Research Council through grant QnanoMECA (CoG - 64790), Fundació Privada Cellex and the Spanish Ministry of Economy and Competitiveness through the Severo Ochoa Programme for Centres of Excellence in R$\&$D (SEV-2015-0522), grant FIS2016-80293-R. 
The author acknowledges useful discussions and related work\citep{conangla2018optimal} with the levitodynamics team of the Plasmon Nano-optics (Romain Quidant) group, at ICFO. 

\end{acknowledgement}


\bibliography{./references}

\newpage
\section{Supplementary material}

\subsection{Linear time-invariant systems}
A linear constant-coefficient ordinary differential equation, defined by a linear polynomial in the unknown function $x(t)$ and its derivatives, takes the form
$$\label{LTI}
f\left(x, x', x'',\ \ldots,\ x^{(n)}\right) = g(t),
$$
where $f(\ldots)$ is a linear function and we added a non-homogeneous term $g(t)$. If we now Fourier transform both sides of the equation, we get
$$
p(i\omega) X(\omega) = G(\omega),
$$
where $p(i\omega)$ is the characteristic polynomial of the differential equation and $X(\omega)$ and $G(\omega)$ are the Fourier transforms of $x(t)$ and $g(t)$ respectively. Therefore
$$
X(\omega) = \frac{1}{p(i\omega)}G(\omega) = H(\omega)\cdot G(\omega),
$$
where we defined the frequency response (or transfer function) $H(\omega) = \frac{1}{p(i\omega)}$. By the convolution theorem, if we apply the inverse Fourier transform we obtain
$$
\mathcal{F}^{-1} \left(X(\omega)\right) = \mathcal{F}^{-1} \left(H(\omega) \cdot G(\omega)\right) $$
$$
x(t) = h(t)*g(t).
$$
Here, $h(t)$ is known as the \emph{impulse response} of the system, and is used to find $x(t)$ for an arbitrary driving $g(t)$.

\subsection{The harmonic oscillator}
The harmonic oscillator is a 2nd order constant coefficient linear ODE that can be used as a first approximation or a paradigmatic model of most simple oscillatory systems. In the most general case, starting from Newton's second law, the equation reads
$$
m\ddot{x} + m\Gamma \dot{x} + kx = F(t)
$$
where $m\Gamma\dot{x}$ is a damping force, $kx$ is a restoring force\footnote{For now we are ignoring the fact that a damping force leads to a Brownian random force, due to the fluctuation-dissipation theorem. Since this force scales with $\sqrt{m}$, the randomness doesn't need to be taken into account for large oscillator masses, but will be relevant in micro and nanoparticles.} (of arbitrary origin) and $F(t)$ a driving force. 

\subsubsection{Conserved quantities}
When $\Gamma = F(t) = 0$, the system is Hamiltonian, and its energy takes the form 
$$
H = \frac{{p}^2}{2m} + \frac{1}{2} m \omega_0^2 {x}^2
$$
where $\omega_0 = \sqrt{\frac{k}{m}}$; it can be proved that this is the only conserved quantity of the system (Hamiltonian systems with $2n$ degrees of freedom can have, at most, $n$ constants of motion). The equality between energy and Hamiltonian is justified because $H$ doesn't explicitly depend on $t$. 

The first summand is the kinetic energy, and the second the (restoring force) potential energy. When feedback is introduced to the system, $H \equiv H(t)$, energy conservation cannot be taken for granted, as in general feedback can lead to cooling, heating, chaotic motion, etc.

\subsubsection{Deterministic solution}
\textbf{Damped harmonic oscillator}: 
If $F(t)$ is zero, we define the \underline{natural frequency} $\omega_0 = \sqrt{\frac{k}{m}}$, the \underline{damping ratio} 
$$
\zeta = \sqrt{\frac{m}{k}}\frac{\Gamma}{2} = \frac{\Gamma}{2 \omega_0}
$$
and the \underline{quality factor} 
$$
Q = \frac{\omega_0}{\Gamma} = \frac{1}{2\zeta}$$
The quality factor can be understood in different (but mostly equivalent) manners. One way, that will be discussed later, is as the number of ``coherent'' oscillations of the system. A different definition is as the fraction of the energy $E$ stored in the system versus the energy dissipated $\Delta E$ in a period $\tau_0$ of the oscillation, as
$$
Q = \frac{2 \pi E}{\Delta E} = \frac{2 \pi}{1 - e^{-\Gamma t_0}} \approx \frac{\omega_0}{\Gamma}.
$$
The last approximation comes from an order one Taylor expansion, so when $\Gamma$ is small one recovers the previous quality factor definition. Using these parameters, the ODE now takes the form
$$
\frac{\mathrm{d}^2x}{\mathrm{d}t^2} + \Gamma\frac{\mathrm{d}x}{\mathrm{d}t} + \omega_0^2 x = 0
$$
The value of the damping ratio $\zeta$ critically determines the behaviour of the system. A damped harmonic oscillator can be 
\begin{itemize}
\item Overdamped, $Q \leq 0.5$. The system exponentially decays to zero without oscillating. The case $Q = 0.5$ is usually called critically damped and is the boundary between oscillation and no oscillation.
\item Underdamped, $Q > 0.5$. The system oscillates at $\omega_1 = \omega_0\sqrt{1 - Q^2/4}$, so the smaller the damping the closer the oscillation frequency to the natural frequency of the oscillator. The sinusoid has an exponential decay of $\lambda = \frac{\omega_0}{2Q} = \frac{\Gamma}{2}$.\footnote{The characteristic time is $\tau = 1/\lambda$}
\end{itemize}

\textbf{Driven harmonic oscillator}:
In the case of a sinusoidal driving force:

$$
\frac{\mathrm{d}^2x}{\mathrm{d}t^2} + \Gamma\frac{\mathrm{d}x}{\mathrm{d}t} + \omega_0^2 x = \frac{1}{m} F_0 \sin(\omega t)
$$

where $F_0$ is the driving amplitude and $\omega$ is the driving frequency for a sinusoidal driving mechanism\footnote{This type of system appears in AC driven inductor-capacitor systems and in driven spring systems having internal mechanical resistance or external, like a particle in an optical trap}.

The general solution is a sum of a transient term that depends on initial conditions, and a steady state that is independent of initial conditions and depends only on the driving amplitude $F_0$, driving frequency $\omega$, undamped angular frequency $\omega_0$, and $\Gamma$. The transient solutions are the same as the unforced ($F_0 = 0$) harmonic oscillator and represent the systems response to other events that occurred previously. However, they typically die out rapidly enough that they can be ignored.

The steady-state solution is proportional to the driving force with an induced phase change of $\phi$:
$$
x(t) = \frac{F_0}{Z_m} \sin(\omega t + \phi)
$$
where
\begin{align*}
Z_m & = m\sqrt{\left(2\omega_0\zeta\right)^2\omega^2 + \left(\omega_0^2  - \omega^2\right)^2} = m\sqrt{\left(\frac{\omega_0}{Q}\right)^2\omega^2 + \left(\omega_0^2  - \omega^2\right)^2} \\
& = m\sqrt{\Gamma^2\omega^2 + \left(\omega_0^2  - \omega^2\right)^2}
\end{align*}
is the absolute value of the linear response function, and
$$
\phi = \arctan\left(\frac{2\omega \omega_0\zeta}{\omega^2-\omega_0^2 }\right)
$$
is the phase of the oscillation relative to the driving force, if the $\arctan(\cdot)$ value is taken to be between -180 degrees and 0 (that is, it represents a phase lag, for both positive and negative values of the $\arctan$ argument).

These last two expressions are obtained from the complex transfer function \label{harmonic_transfer}
\begin{align*}
H(\omega) = \frac{1}{m(\omega_0^2 - \omega^2 + i\omega \Gamma)}, 
\end{align*}
the power transfer function being
$$
|H(\omega)|^2 = \frac{1/m^2}{\Gamma^2\omega^2 + \left(\omega_0^2  - \omega^2\right)^2}
$$

For a particular driving frequency, called the resonance or resonant frequency 
$$
\omega_r = \omega_0\sqrt{1-2\zeta^2} = \omega_0\sqrt{1-\frac{1}{2Q^2}}
$$
the amplitude (for a given $F_0$) is maximum, and
$$
|H(\omega_r)|^2 = \frac{4Q^4}{m^2(4Q^2-1)\omega_0^4} \simeq \frac{Q^2}{m^2\omega_0^4}.
$$
The last approximation only holds for large $Q$ factors; notice that the ratio between the power transfer function at $\omega_r$ and at zero is $H(\omega_r)^2/H(0)^2 \simeq Q^2$. 

The resonance effect only occurs when $Q > \sqrt{2}/2$, i.e. for significantly underdamped systems. For strongly underdamped systems the value of the amplitude can become quite large near the resonance frequency.

We can calculate the half width half maximum (HWHM) by imposing $Z_m^2$ to be half of the value at resonance. The angular frequencies thus obtained are
$$
\omega_i = \sqrt{\omega_r^2 \pm \frac{\omega_0^2}{Q^2}\sqrt{Q^2 - \frac{1}{4}}}
$$
and by approximating $\sqrt{Q^2 - \frac{1}{4}} \simeq Q$, $\omega_r \simeq \omega_0$ and applying the Taylor series expansion of the square root in $\omega_0\sqrt{1 \pm \frac{1}{Q}} $ , we obtain
$$
\text{HWHM} = \frac{\omega_2 - \omega_1}{2} \simeq \frac{\omega_0}{2}\left(1 + \frac{1}{2Q} - \frac{1}{8Q^2} - 1 + \frac{1}{2Q} + \frac{1}{8Q^2} + \mathcal{O}\left(\frac{1}{Q^3}\right)\right) = \frac{\omega_0}{2Q} = \frac{\Gamma}{2}
$$
This shows that increasing $Q$ (or, equivalently, decreasing $\Gamma$) reduces the width of the frequency response of the oscillator. Therefore, a high $Q$ is particularly important when trying to detect frequency shifts.

\subsubsection{Stochastic driving forces}
If the intrinsic (classical) randomness of the system is taken into account via the fluctuation-dissipation theorem, then the driving force will have a stochastic driving term $F(t) = \sigma \eta(t)$, with $\eta(t)$ a zero mean and unit standard deviation white noise\footnote{This is not exactly true, as white noise doesn't exist; a proper treatment requires the use of Ito calculus. However, it is still useful to think of it as white noise.} and $\sigma = \sqrt{2 k_B T \gamma}$  (obtained from a fluctuation-dissipation theorem). The full system can be solved in all generality, but it is useful to consider the ``overdamped'' regime first.

\textbf{Overdamped regime}:
It is common to discard the second order term of the stochastic differential equation when it is ``small'' compared to the other terms\footnote{This can be made rigorous as a perturbative problem, a \emph{regular perturbation} problem. However, this is not always the case when the parameter is multiplying the highest order term of the equation. See ``Singular perturbation problem'' for more details}. In this case the equation takes the form of an Ornstein-Uhlenbeck process
$$
\gamma \dot{x} + \delta x = \sigma \eta(t)
$$
with $\gamma$ the damping constant and $\delta = m\omega_0^2$ the restoring force. In Ito's notation
$$
\dif X_t = -a X_t \,\dif t + b\, \dif W_t
$$
with $a = \frac{\delta}{\gamma} > 0$, $b = \frac{\sigma}{\gamma}$.

The solution of this SDE is
$$
X_t = X_0 e^{-at} + b \int_0^t e^{-a(t-s)} \dif W_s.
$$
To see how the process will diffuse, we can calculate its variance process with Ito's isometry:
$$
\mathbb{E}(X_t^2) = \mathbb{E} \left[b \int_0^t e^{-a(t-s)}\,\dif W_s\right]^2 = b^2 \mathbb{E} \left[ \int_0^t \left(e^{-a(t-s)}\right)^2\,\dif s\right]^2 = \frac{b^2}{2a}\left( e^{-2at} - 1 \right),
$$
which, for very short times (i.e., applying a first order Taylor expansion), scales as 
$$
\mathbb{E}(X_t^2) \simeq b^2 t = \frac{\sigma^2}{\gamma^2}t.
$$

The autocorrelation can also be obtained by using Ito's isometry on the last expression
\begin{align*}
R(t,s) & = \mathbb{E}  \left[ \int_0^t f(u)\,\dif W_u \int_0^s f(v) \,\dif W_v \right] \\
& =  b^2 e^{-a(s+t)}\cdot \mathbb{E} \left[ \int_0^t e^{au}\,\dif W_u \int_0^s e^{av} \,\dif W_v \right] 
= b^2 e^{-a(s+t)}\cdot  \mathbb{E} \left[ \int_0^{\min(t,s)}e^{au}e^{au} \,\dif u \right] \\
& = \frac{b^2}{2a} e^{-a(s+t)}(e^{2\min(s,t)} - 1) = \frac{b^2}{2a} \left( e^{-a|t-s|} - e^{-a(t+s)}\right).
\end{align*}

When $|t-s| = 0$, the autocorrelation equals the variance of $X_t$,\footnote{Equivalently this can also be written as $\langle X_t^2 \rangle$}
$$
\mathbb{E}(X_t^2) = \frac{b^2}{2a}\left(1 - e^{-2at}\right) = \frac{k_B T}{m\omega_0^2}\left(1 - e^{-2at}\right).
$$
For small $t$, the variance takes the expression
$$
\lim_{t \rightarrow 0} \mathbb{E}(X_t^2) = \frac{2 k_B T}{\gamma}t + \mathcal{O}(t^2) = \frac{\sigma^2}{\gamma^2}t + \mathcal{O}(t^2),
$$
while for long times
$$
\lim_{t \rightarrow \infty} \mathbb{E}(X_t^2) = \frac{k_B T}{m\omega_0^2},
$$
as is expected from the equipartition theorem. As $t$ and $s$ increase, the second exponential summand of the process autocorrelation becomes arbitrarily small. The remaining part is a function of $\tau \equiv t-s$ only; thus, we can apply the Wiener-Khinchin theorem to get an analytical expression of the power spectrum of the process:
$$
S(f) = \mathcal{F}\left(\frac{b^2}{2a} e^{-a|\tau|}\right) = \frac{b^2}{a^2 + 4\pi^2 f^2}
$$
\begin{align}
S(\omega) = \frac{\sigma^2}{\delta^2 + \gamma^2 \omega^2} = \frac{\sigma^2/m^2}{\omega_0^4 + \Gamma^2 \omega^2}
\end{align}
where in the last expression we have reintroduced the mass, as it will help in identifying similarities between this and the full second order system power spectrum. 

As expected by Parseval's theorem, the integral of the power spectral density is
\begin{align*}
\mathbb{E}(X_t^2) = \frac{1}{2\pi} \int S(\omega) \dif \omega = \frac{1}{2\pi} \int \frac{\sigma^2/m^2}{\omega_0^4 + \Gamma^2 \omega^2} \dif \omega = \frac{k_B T}{m \omega_0^2}
\end{align*}
recovering again the value expected by the equipartition theorem.

\begin{center}
\fbox{
\begin{minipage}{0.9\textwidth}
\textbf{Observation}: The expression of the power spectral density has two clear different regimes: at low frequencies, the $a^2$ term in the denominator dominates and the spectrum is almost flat. At large $f$, $a^2$ is negligible and the other term dominates. In a log-log scale representation, the power spectrum looks like two straight lines (first an horizontal line and then a decreasing line); the frequency at which the behaviour changes is known as the \emph{corner frequency}, $\omega_c = a = \frac{\delta}{\gamma} = \frac{\omega_0^2}{\Gamma}$.
\end{minipage}
}\end{center}

\textbf{Full 2nd order equation}:
The full equation takes the expression
$$
m \ddot{x} + \gamma \dot{x} + k x = \sigma \eta(t)
$$
Performing a change of variables, $x = e^{\frac{-\gamma}{2m}t}x_1$, we get
$$
m \ddot{x}_1 + \left(k - \frac{\gamma^2}{4m}\right)x_1 = \sigma e^{\frac{\gamma}{2m}t}\eta(t)
$$
thus eliminating $\dot{x}$. Setting $a = \frac{k}{m} - \frac{\gamma^2}{4m^2}$, $b = \frac{\sigma}{m}$ and rewriting the equation as a first order linear system with
$$
X = 
\begin{pmatrix}
x_2\\ 
v_2
\end{pmatrix},
$$
where $v_2 = \dot{x_2}$, we get in Ito's notation
\begin{equation}
\dif X = 
\begin{pmatrix}\label{eq:complete}
0 & 1 \\ 
-a & 0
\end{pmatrix}
\cdot X \dif t +
\begin{pmatrix}
0\\ 
b e^{\frac{\gamma}{2 m}t}
\end{pmatrix}
\cdot \dif W_t
\end{equation}
The solution of a linear homogeneous SDE is
$$
X_t = e^{\int_0^t A(t)\dif t}\cdot X_0 + e^{\int_0^t A(t)\dif t}\cdot \int_0^t e^{-\int A(s)\dif s}\sigma(s)\dif W_s
$$
where $A(t)$ is the (generally vector) coefficient of $X$. For this SDE a fundamental matrix solution of the associated homogeneous noise-free system is
$$
\Phi(t) = 
\begin{pmatrix}
\cos \sqrt{a}t & \sin \sqrt{a}t/\sqrt{a} \\ 
-\sqrt{a} \sin\sqrt{a}t & \cos \sqrt{a}t
\end{pmatrix}
$$
The determinant of this matrix is 1, so its inverse matrix will be
\small
$$
\Phi^{-1}(t) = e^{-\int A(\tau)\dif \tau} = \det \Phi(t)^{-1}\cdot
\begin{pmatrix}
\cos \sqrt{a}t & -\sin \sqrt{a}t/\sqrt{a} \\ 
\sqrt{a} \sin\sqrt{a}t & \cos \sqrt{a}t
\end{pmatrix}
=
\begin{pmatrix}
\cos \sqrt{a}t & -\sin \sqrt{a}t/\sqrt{a} \\ 
\sqrt{a} \sin\sqrt{a}t & \cos \sqrt{a}t
\end{pmatrix}
$$
\normalsize
and hence we can solve the complete system. We are interested in the first component of $X$, the position (as we will be calculating the PSD of the trajectory of the particle)

\footnotesize
\begin{equation*}
\begin{split}
x_1(t) = &
\begin{pmatrix}
\cos \sqrt{a}t  & \sin \sqrt{a}t/\sqrt{a}
\end{pmatrix}
\cdot
\begin{pmatrix}
x_1(0)\\ 
v_1(0)
\end{pmatrix} 
+
\begin{pmatrix}
\cos \sqrt{a}t  & \sin \sqrt{a}t/\sqrt{a}
\end{pmatrix}
\cdot
\int_0^{t}
b e^{\frac{\gamma}{2m}r}\cdot
\begin{pmatrix}
-\sin \sqrt{a}r/\sqrt{a}\\ 
\cos \sqrt{a}r
\end{pmatrix}
\dif W_r
\end{split}
\end{equation*}
\normalsize
Finally, $x_1(t) = e^{\frac{\gamma t}{2m}}x(t)$, so
\footnotesize
\begin{equation}
\begin{split}\label{stochastic_solution}
x(t) = &
e^{-\frac{\gamma t}{2m}}\begin{pmatrix}
\cos \sqrt{a}t  & \sin \sqrt{a}t/\sqrt{a}
\end{pmatrix}
\cdot
\begin{pmatrix}
x(0)\\ 
v(0) + \frac{\gamma}{2m}x(0)
\end{pmatrix} 
+\\
& 
e^{-\frac{\gamma t}{2m}}
\begin{pmatrix}
\cos \sqrt{a}t  & \sin \sqrt{a}t/\sqrt{a}
\end{pmatrix}
\cdot
\int_0^{t}
b e^{\frac{\gamma}{2m}r}\cdot
\begin{pmatrix}
-\sin \sqrt{a}r/\sqrt{a}\\ 
\cos \sqrt{a}r
\end{pmatrix}
\dif W_r
\end{split}
\end{equation}
\normalsize

We see that, after a transient time, only the term depending on $\dif W_r$ remains, so the first moment of the process is zero. Now, applying Ito's isometry as before to calculate the covariance we get
\footnotesize
$$
R(t,s) = \mathbb{E}  \left[ \int_0^{t} f(u)\,\dif W_u \int_0^s f(v) \,\dif W_v \right] =
$$
\begin{align*}
b^2 e^{-\frac{\gamma (t + s)}{2m}}
\begin{pmatrix}
\cos \sqrt{a}t  & \sin \sqrt{a}t/\sqrt{a}
\end{pmatrix}
& \cdot \mathbb{E} \left[ \int_0^{\min(t,s)} 
e^{\frac{\gamma}{m}u}
\begin{pmatrix}
\frac{\sin^2 \sqrt{a}u}{a}  & -\frac{\sin \sqrt{a}u \cos \sqrt{a}u}{\sqrt{a}}\\
-\frac{\sin \sqrt{a}u \cos \sqrt{a}u}{\sqrt{a}} & \cos^2 \sqrt{a}u
\end{pmatrix}
\, \dif u \right]
\cdot \\
& \cdot
\begin{pmatrix}
\cos \sqrt{a}s  \\
\sin \sqrt{a}s/\sqrt{a}
\end{pmatrix}.
\end{align*}
\normalsize
This is a quite uninteresting calculation\footnote{The results are verified with Mathematica. I have the file in ./Simulations/Mathematica}. As in the one-dimensional Ornstein-Uhlenbeck process, one gets a term which only depends on the difference $|t - s| = \tau$, and another which is multiplied by $e^{-|t + s|}$, that quickly vanishes. Setting $t = s$ we recover the variance,
\begin{align}
\mathbb{E}[X^2_t] = \frac{k_B T}{m \omega_0^2}\left(1 - e^{-\Gamma t}\left(\frac{\omega_0^2}{{\omega_1^2}} -  \frac{\Gamma^2}{4\omega_1^2} \cos(2\omega_1 t) + \frac{\Gamma}{2\omega_1} \sin(2\omega_1 t)\right)\right)
\end{align}
where we have defined $\omega_1 = \sqrt{\omega_0^2 - (\Gamma/2)^2} = \sqrt{a}$ and used the normalized damping constant $\Gamma = \frac{\gamma}{m}$. As in the overdamped case,
$$
\lim_{t \rightarrow \infty} \mathbb{E}[X^2_t] = \frac{k_B T}{m \omega_0^2},
$$
as expected from the equipartition theorem. However, more interesting is the behaviour at short times, 
$$
\lim_{t \rightarrow 0} \mathbb{E}[X^2_t] = \frac{2\Gamma k_B T}{3m}t^3 + \mathcal{O}(t^4).
$$

\begin{center}
\fbox{
\begin{minipage}{0.9\textwidth}
\textbf{Observation}: this value of the variance assumes that $x_0$, $v_0$, initial conditions of the problem, are known. This \emph{may not be the case} for a real experiment in the lab. For instance, assume we have a particle in an optical tweezer, and we want to calculate the mean square displacement (MSD), 
$$
\mathbb{E}[(x(t) - x(0))^2] = \mathbb{E}[x^2(t) + x^2(0) - 2x(0)x(t)].
$$ 
Then we distinguish two cases:
\begin{enumerate}
\item $x(0)$ is known, e.g. $x(0) = 0$. Then
$$ 
\mathbb{E}[x^2(t) + x^2(0) - 2x(0)x(t)] = \mathbb{E}[x^2(t)] = \mathcal{O}(t^3).
$$
\item $x(0)$ is not known (or is uncontrolled), and we do the ensemble average. Then
\begin{align*}
\mathbb{E}[x^2(t) + x^2(0) - 2x(0)x(t)] & = \mathbb{E}[x^2(t)] + \mathbb{E}[x^2(0)] - 2\mathbb{E}[x(0)x(t)] \\
& = \frac{2 k_B T}{m \omega_0^2} - 2R(t) = \mathcal{O}(t^2)
\end{align*}
\end{enumerate}
We see, therefore, that the leading term of the series expansion of the MSD is different in each case. This is an important distinction that is not clear in Toncang's Li Science paper\footnote{R. Rica contributed to this.}
\end{minipage}
}\end{center}

If in the autocorrelation integral we keep only the term with $\tau$ dependency (since the term multiplied by $e^{-|t+s|}$ will quickly decay), after some simplification one gets
$$
R(\tau) = \frac{b^2 m^2 e^{\frac{-\Gamma |\tau|}{2}}\left( 2 \sqrt{a}m \cos(\sqrt{a}|\tau|) + \gamma \sin(\sqrt{a}|\tau|) \right)}{\sqrt{a}(\gamma^3 + 4a\gamma m^2)}.
$$
Worth mentioning is the value of the time constant of the exponential factor, $\frac{2}{\Gamma}$: this value is proportional to the time needed for the autocorrelation to get below a given threshold\footnote{Which is arbitrarily set}, otherwise known as \emph{losing the coherence} of the oscillation. A usual criterion is 3 time constants, with the autocorrelation decreasing to below $e^{-3} < 5\%$. The number of oscillations of the system during this time is
$$
3\frac{2/\Gamma}{1/f} = \frac{3\cdot 2 \cdot f}{\Gamma} \simeq \frac{\omega}{\Gamma} = Q
$$
Thus, the quality factor can be understood as the number of oscillations of the system under the presence of Brownian noise before the autocorrelation gets below 5$\%$\footnote{The number of ``coherent'' oscillations}.

From the expression of the autocorrelation we see that $R(t,\tau) = R(\tau)$: therefore, the process is wide-sense stationary and the conditions to apply the Wiener-Khinchin theorem are satisfied. The Fourier transform of this autocorrelation function is the power spectral density 
$$
S(f) = \frac{16b^2}{\Gamma^4 + 8(a + 4\pi^2f^2)\Gamma^2 + 16(a-4\pi^2f^2)^2}
$$
which, after replacing the variables and some rearranging\footnote{See, again, the Mathematica file ``mathem\_psd\_calculations''} takes the simpler and more familiar expression
$$
S(\omega) = \frac{\sigma^2/m^2}{(\omega_0^2 - \omega^2)^2 + \Gamma^2 \omega^2}
$$
where we have replaced the unitary ordinary frequency Fourier transform (in terms of $f$) by the non-unitary angular frequency Fourier transform. Thus, it is readily seen that this is the frequency response of a harmonic oscillator to a flat-spectrum (white noise) driving force.

As in the overdamped case, 
$$
\mathbb{E}(X_t^2) = \frac{1}{2\pi} \int \frac{\sigma^2/m^2}{(\omega_0^2 - \omega^2)^2 + \Gamma^2 \omega^2} \dif \omega = \frac{k_B T}{m \omega_0^2}
$$
as expected by the equipartition theorem.
\end{document}